\documentclass[a4paper,12pt,english]{article}

\usepackage{amsfonts,bm,amssymb,euscript,array,babel,cite}
\usepackage[all,color]{xy}
\usepackage{relsize}
\usepackage{url}
\usepackage{amsmath,amsthm,accents,bm}
\usepackage[]{graphicx}
\usepackage[T1]{fontenc}
\usepackage{framed}

\usepackage{empheq}
\usepackage{color}
\usepackage{mathtools}
\usepackage{titlesec}

\makeatletter

\newcommand{\dd}{\mathrm{d}}

\newcommand{\w}{\wedge}
\newcommand{\bbm}{\left(\begin{matrix}}
\newcommand{\ebm}{\end{matrix}\right)}
\newcommand{\beq}{\begin{eqnarray}}
\newcommand{\eeq}{\end{eqnarray}}

\makeatother

\newcommand{\sfrac}[2]{{\textstyle\frac{#1}{#2}}}

\newcommand{\be}{\begin{equation}}
\newcommand{\ee}{\end{equation}}

\newcommand{\beqa}{\begin{eqnarray}}
\newcommand{\eeqa}{\end{eqnarray}} 
\def\nn{\nonumber} \def \bea{\begin{eqnarray}} \def\eea{\end{eqnarray}}

\newcommand{\barr}{\begin{array}}
\newcommand{\earr}{\end{array}}
\numberwithin{equation}{section}

\def\a{\alpha}  
  \def\G{\Gamma}
 \def\d{\delta} 
 
 \def\vare{\varepsilon}
    
\def\l{\lambda}   \def\m{\mu}
\def\n{\nu} \def\o{\omega}

\def\mc{\mathcal}

\def\Z{{\mathbb Z}} \def\one{\mbox{1 \kern-.59em {\rm l}}}

\def\bi{\begin{itemize}} \def\ei{\end{itemize}}

\def\tT{\widetilde{\bm {\o}}}
\def\tn{\widetilde{\bm\nabla}}
\def\n2{\bm\nabla^2}

\def\td{\widetilde{\bm\dd}}

\begin{document}

\makeatother

\parindent=0cm

\renewcommand{\title}[1]{\vspace{10mm}\noindent{\Large{\bf

#1}}\vspace{8mm}} \newcommand{\authors}[1]{\noindent{\large

#1}\vspace{5mm}} \newcommand{\address}[1]{{\itshape #1\vspace{2mm}}}

\begin{titlepage}

\begin{flushright}
\today \\
\end{flushright}

\begin{center}

\vskip 3mm

\title{ {\LARGE
Tensor Galileons and Gravity
} }

 \authors{Athanasios {Chatzistavrakidis}$^{\a}$, 
 Fech Scen Khoo$^{\beta}$, \\ \smallskip Diederik Roest$^{\a}$ and 
 Peter Schupp$^{\beta}$}

\vskip 1mm

\address{
$^{\a}$Van Swinderen Institute for Particle Physics and 
Gravity, University of Groningen, \\ 
Nijenborgh 4, 9747 AG Groningen, The Netherlands

\smallskip 

$^{\beta}$Department of Physics and Earth Sciences, Jacobs University Bremen, \\ 
Campus Ring 1, 28759 Bremen, Germany

}

\smallskip

\verb+a.chatzistavrakidis@rug.nl, f.khoo@jacobs-university.de,+ \\ 
\verb+d.roest@rug.nl, p.schupp@jacobs-university.de+

\end{center}

\vskip 1mm

 \begin{center}
\textbf{Abstract}
\vskip 3mm
\begin{minipage}{14cm}%
The particular structure of Galileon interactions allows for higher-derivative terms while retaining second order field equations for scalar fields and Abelian $p$-forms. In this work we introduce an index-free formulation of these interactions in terms of two sets of Grassmannian variables. We employ this to construct Galileon interactions for mixed-symmetry tensor fields and coupled systems thereof. {We argue that these tensors are the natural generalization of scalars with Galileon symmetry, similar to $p$-forms and scalars with a shift-symmetry.} The simplest case corresponds to  linearised 
{gravity} with Lovelock invariants, {relating the Galileon symmetry to diffeomorphisms.} Finally, we examine the coupling of a mixed-symmetry tensor to gravity, and demonstrate in an explicit example that the inclusion of appropriate counterterms retains second order field equations.

\end{minipage}

 \end{center}

\end{titlepage}

\tableofcontents

\newpage

\setcounter{footnote}{0}

\section{Introduction}
\label{intro}

When does an action functional lead to field equations which contain only second derivatives of the 
corresponding field? Given that second order field equations are very popular in Nature, largely due to the lack of additional ghost-like degrees of freedom \cite{Ostrogradsky} that generically appear in higher order field equations\footnote{Remarkably, it has recently been found that one can have ghost-free higher-order field equations in coupled systems, such as tensor-scalar theories \cite{Gleyzes:2014dya} and classical mechanics \cite{Motohashi:2016ftl}; we will restrict ourselves to second order field equations, however.}, the above question has always drawn considerable interest. 

In the case of a rank-2 symmetric tensor, namely a metric, this question was posed and answered long ago in any dimension, leading to the Lovelock invariants \cite{lovelock}. Similarly, for  scalar fields this question is answered by Galileons \cite{Nicolis:2008in,generalizedgalileons}, the name reflecting the property that due to these second derivatives there is a Galilean-type of symmetry for the scalar fields in the equations of motion. More recently, a number of generalizations  have been suggested. These include scalar theories with field equations containing derivatives up to second 
order instead of strictly second order \cite{ggalileons}. Secondly, these theories have been coupled to gravity, leading to covariant Galileons \cite{covariantgalileons}. In four dimensions, this leads to the most general scalar-tensor theory constructed by Horndeski  \cite{horndeski}. Finally, Galileon interactions have been constructed for other types of 
fields, such as vectors and $p$-forms \cite{pformgalileons,pformgalileons2,pformgalileons3} and combinations of fields of 
different type. These constructions are reviewed in Ref.~\cite{reviewgalileons}. 

{\it With the exception of Lovelock, all of these Galileon interactions concern fields with at most spin-1. Is a similar structure possible for higher-spin fields as well? We will answer this question affirmatively for all spin-2 fields, and demonstrate that   gravity's diffeomorphisms are actually closely related to the Galileon symmetry.}

Mixed-symmetry tensor fields are covariant tensors with more than one set of 
antisymmetrized indices, relevant in particular for spin $\geq 2$. Recalling that $p$-forms are fully antisymmetric covariant tensor fields, 
mixed-symmetry tensors are a natural generalization of differential forms, which neither need to be fully antisymmetric nor fully symmetric. 
From a different point of view, 
they correspond to Young diagrams with more than one column. Such tensor fields were studied by Curtright
in \cite{curtright} in the context of  generalizations of gauge theory. From a more modern 
viewpoint, interest in mixed-symmetry tensor fields was sparked due to their appearance in string theory and 
in studies of the dual graviton. For example, notably such fields appear naturally in $E_{11}$ \cite{West1}, 
but also as fields coupling to non-standard branes in string theory 
\cite{Bergshoeff3}, and as exotic duals to standard 
fields \cite{demedeiroshull1,DuboisViolette:2001jk}.

The goal of this paper is twofold. Our first objective is to formulate Galileon action functionals in an elegant and index-free formalism, rather than the usual formulation which can be {complicated on account of the many indices.} 
Let us recall that the action for the scalar Galileon $\pi$, occuring $n+1$ times in $D\ge n$ dimensional flat space, can be written as\footnote{This is not the only possible form of the action for scalar Galileons, since one may wish to perform additional partial integrations. 
Other suggested forms are presented for instance in Ref. \cite{reviewgalileons}.}  
\bea \label{pigal}
S^{\text{Gal,1}}_{n+1}[\pi]&=&\int\dd^D x\, {\cal A}_{(2n)}^{i_1\dots i_{n}j_1\dots j_{n}}
\partial_{i_1}\pi\partial_{j_1}\pi\partial_{i_2}\partial_{j_2}\pi\dots\partial_{i_{n}}
\partial_{j_{n}}\pi~,
\eea
where 
\be \label{pigal2}
{\cal A}_{(2n)}^{i_1\dots i_{n}j_1\dots j_{n}}=\frac{1}{(D-n)!}\vare^{i_1\dots i_{n}k_1\dots k_{D-n}}
\vare^{j_1\dots j_{n}}{}_{k_1\dots k_{D-n}}~,
\ee
and $\vare$ is the totally antisymmetric epsilon symbol in $D$ dimensions. The reason for the name Galileon is the invariance of this 
action under the symmetry $\partial_{i}\pi\to \partial_i\pi+b_i$ and $\pi \to \pi + b_i x^i+c$, for constant $c$ and $b_i$.
 Similar action functionals may be 
written for differential forms of any even degree, and also for coupled systems of scalar fields and differential forms of any degree 
\cite{pformgalileons}. 

Noting that the key structure that allows for the 
construction of Galileons is the presence of two separate fully antisymmetric {Levi-Civita tensors},
we suggest that a natural formalism to account for this feature is in terms of graded, anticommuting variables.
Thus in Section \ref{gggal} we introduce two sets of such variables and also accommodate the degrees of freedom, here 
$p$-forms, in quantities constructed with the aid of 
these anticommuting variables. Differential operators acting on the fields are also constructed 
accordingly. This results in a formulation where the Levi-Civita tensors are replaced by (Berezin) integration over 
the graded variables, no indices appear, and a compact yet full (meaning non-formal) expression is possible\footnote{An interesting
formulation of scalar-tensor 
theories in the language of differential forms was suggested in Ref. \cite{Ezquiaga:2016nqo}. 
}. 
For instance, we will show that the scalar Galileon action \eqref{pigal} with \eqref{pigal2} may be expressed as 
\begin{empheq}[box=\fbox]{align}
S^{\text{Gal,1}}_{n+1}[\pi] = \frac 1{(D-n)!}\int\dd^Dx\int d^D\theta \, d^D  \chi \, \bm{\eta}^{D-n} \,
\bm{\dd} \pi \, \widetilde{\bm{\dd}}\pi \,  (\bm{\dd}\widetilde{\bm{\dd}} \pi )^{n-1}~, 
\label{ggscalar-intro}
\end{empheq}
where the quantities appearing in this action will be explained in the main text. 

Our second objective is to generalize these action functionals for mixed-symmetry tensor fields. 
First we consider a single $(p,q)$ tensor field, meaning that out of its $p+q$ covariant 
indices, $p$ and $q$ of them are separately antisymmetrized. The action we find has a similar structure to 
\eqref{pigal}, 
and leads to strictly second order field equations for the $(p,q)$ tensor field in flat 
space-time. This requirement is the guiding principle for our analysis. This is different in spirit to \cite{curtright}, where the starting point was a postulated gauge invariance. The actions we find do have the corresponding gauge symmetries, but we see this as a consequence and not as a starting point. In fact, there are also symmetries that are not of gauge-type.
Using our formulation in terms of graded variables, we further 
generalize the mixed-symmetry Galileons in the  directions of (i) requiring field 
equations up to second order, and (ii)  combining arbitrary number of mixed-symmetry tensor fields 
of arbitrary degree in a coupled system.  We provide some  examples to illustrate these generalizations.
Notably, in the simplest case of a $(1,1)$ tensor field, being a rank-2 symmetric tensor, the resulting single-field action is identified with the linearised Lovelock invariants in any dimension. 

{\it An interesting observation that follows from our analysis is that these $(p,q)$ mixed-symmetry fields can be seen as the natural tensor generalization of scalar fields with Galileon symmetry. This is akin to the $p$-form generalization of scalar fields with a shift symmetry, i.e.~which are only derivatively coupled.}

Finally, we consider covariant mixed-symmetry Galileons, namely the curved space-time versions of the action 
functionals for mixed-symmetry tensor fields. This is interesting because it introduces higher derivative couplings between 
an external metric and the spin-2 fields we study. There are two issues regarding this type of theories. 
One is the question whether the property of second order field equations can remain intact upon introduction of 
counter-terms in the action. The second is whether the resulting theories are consistent,
even when they are safe from the Ostrogradsky ghost. It is known that 
for a large class of cases and under {rather} general assumptions, the latter is hard to achieve 
\cite{Boulanger:2000rq,Aragone:1979bm}. 
Here we argue that already the first question is not trivial, and is worth investigating. In particular, we show that 
starting with a 5D, 3-field Galileon for the (1,1) mixed-symmetry tensor, it is indeed possible to find compensating 
terms, which retain the second order nature of the field equations. We believe that the remarkable cancellations occurring in this particular case may extend to general mixed-symmetry Galileons coupled to gravity.

\section{Galileons and graded variables}
\label{gggal}

\subsection{Preliminaries on anticommuting variables}
\label{gggal1}

Given that the actions for Galileons are highly non-linear, they appear rather complicated and employ a large number of indices. Thus we find it useful to 
suggest an index-free formulation, which simplifies the notation and computations and furthermore renders the 
properties of the actions more transparent and easier to generalize. 
Given that we are dealing with more than one set of antisymmetrized indices, as explained in the Introduction,
it is convenient to introduce two sets of anticommuting, graded variables 
$(\theta^{i})$ and $(\chi^{i})$. Extending the bosonic (degree-0) coordinates $(x^i)$ by these degree-1 variables essentially means that 
we are going to write down actions on a graded (super)manifold\footnote{Despite the resemblance to supersymmetry, 
this does not mean that the theories we consider are supersymmetric. The role of graded geometry here is that of a tool to formulate
the bosonic actions in an elegant 
fashion. This should be also clear from the fact that we use the same type of indices to count the anticommuting 
coordinates. Supersymmetric Galileons were considered in \cite{Khoury:2011da}.} ${\cal M}$. 
These variables satisfy the following properties:
\be 
\theta^i\theta^j=-\theta^j\theta^i~, \quad \chi^i\chi^j=-\chi^j\chi^i~, \quad \theta^i\chi^j=\chi^j\theta^i~.
\ee
The first two properties are the standard Grassmann relations, while the 
last property simply means that the two sets of additional variables are mutually commuting. The latter property is conventional, in the sense that 
one could have as well chosen mutually anticommuting variables, which would have led to some  adjustable signs in the formulae. 

Using the degree-1 variables $\theta^i$ and $\chi^i$, a $p$-form $\o^{(p)}$  may be represented as 
\be \label{omegatheta}
\bm\omega^{(p)}=\sfrac 1{p!}\omega_{i_1\dots i_p}\theta^{i_1}\dots\theta^{i_p}~.
\ee
(Henceforth we use boldface notation to denote objects that depend on anticommuting variables.)
Alternatively there is a second such object with the same components,
\be \label{omegachi}
\widetilde{\bm{\omega}}^{(p)}=\sfrac 1{p!}\omega_{i_1\dots i_p}\chi^{i_1}\dots\chi^{i_p}~,
\ee
associated to the second set of graded variables. 
The difference to the standard formalism for $p$-forms is that antisymmetrizations are now controlled by the appearance of the graded coordinates in the corresponding fields.
Evidently, for the degenerate case of 0-forms (scalar fields) it holds that 
$\bm{\omega}^{(0)}=\widetilde{\bm{\omega}}^{(0)}=\omega^{(0)}$, since no graded variables are involved. 
For later convenience, let us also introduce the following objects with the components of the flat metric $\eta_{ij}$ or the curved metric $g_{ij}$ respectively, 
\be 
\bm{\eta}=\eta_{ij}\theta^i\chi^j~, \quad \bm{g}=g_{ij}\theta^i\chi^j~.
\ee
The other important ingredient for our purposes are exterior derivatives. One can define two separate ones, in particular
\be 
\bm{\dd}=\theta^i\partial_i \quad \text{and} \quad \widetilde{\bm{\dd}}=\chi^i\partial_i~.
\ee
Both derivatives are nilpotent,
\be 
\bm{\dd}^2=\theta^i\theta^j\partial_i\partial_j=0 \quad \text{and} \quad {\widetilde{\bm{\dd}}}^2=\chi^i\chi^j\partial_i\partial_j=0~,
\ee
provided that the bosonic  variables $x^i$ and the degree-1 variables $\theta^i$ and $\chi^i$ are independent, which is 
indeed the case here. 
In addition, the two derivatives commute
\be 
\bm{\dd}\widetilde{\bm{\dd}}=\widetilde{\bm{\dd}}\bm{\dd}~, 
\ee
as a consequence of our convention of mutually commuting graded coordinates.

Finally, since no $\theta$s or $\chi$s should appear in the explicit form of the actions we are going to consider, one has to integrate over them. 
Let us recall the basic formula for the Berezin integral,
\be 
\int \dd \theta \, \theta=1
\ee
up to a choice of normalization.
This implies that in $D$ dimensions
\be \label{integral}
\int d^D\theta \, \theta^{i_1} \ldots \theta^{i_D} = \varepsilon^{i_1 \ldots i_D} \,.
\ee
This expression is valid in flat space and with the indices referring to the Minkowski coordinate basis. In curved space and for more general choices of coordinates, the epsilon symbol is to  be replaced by the epsilon tensor, i.e.\ the value of integrals like \eqref{integral} can depend on spacetime position.  
These integral formulas are pivotal in our formulation, since the Lagrangians will always involve integration over the graded variables.

\subsection{Scalar and $p$-form Galileons in the graded formalism}
\label{gggal2}

Let us now use the above formalism to express the actions for scalar Galileons in flat space-time. The curved case will be 
examined in the penultimate section. 
The action for scalar Galileons in $D\ge n$ dimensions with $n+1$ occurences of the scalar field $\pi$ may be written as 
\bea
S^{\text{Gal,1}}_{n+1}[\pi] &=& \frac 1{(D-n)!}\int\dd^Dx\int d^D\theta \, d^D  \chi \, \bm{\eta}^{D-n} \,
\bm{\dd} \pi \, \widetilde{\bm{\dd}}\pi \, (\bm{\dd}\widetilde{\bm{\dd}} \pi )^{n-1} \nn\\
&=& -\frac 1{(D-n)!}\int\dd^Dx\int d^D\theta \, d^D  \chi \, \bm{\eta}^{D-n} \,
 \pi \,  (\bm{\dd}\widetilde{\bm{\dd}} \pi )^{n}
\label{ggscalar}
\eea
up to a boundary term, 
where $\dd^D\theta=\dd\theta^1\dots\dd\theta^D$ and $\dd^D\chi=\dd\chi^1\dots\dd\chi^D$.  The desired property of second order field equations 
is rather evident. 
Indeed, using the identities $\bm{\dd}\bm{\dd}\pi=\widetilde{\bm{\dd}}\widetilde{\bm{\dd}}\pi=0$, we directly obtain that the variation of this action, i.e.~the Euler-Lagrange equation, reads
\be 
E_{n+1}=-\frac{n+1}{(D-n)!}\int\dd^D\theta \, \dd^D\chi \, \bm{\eta}^{D-n} (\bm{\dd}\widetilde{\bm{\dd}}\pi)^n =0~,
\ee
which is obviously second order in derivatives. Higher derivatives of the fields cannot occur, because such expressions would contain $\bm\dd^2$ or $\widetilde{\bm\dd}{}^2$ and would thus be equal to zero.
Despite its simplicity \eqref{ggscalar} is equivalent to \eqref{pigal} as can be seen by evaluating the integrals over the graded variables (see Appendix A).

The generalization of Galileons for higher-degree differential forms and coupled systems thereof is straightforward in the index-free formalism presented here. 
For example, for an Abelian 1-form ${\omega}^{(1)}:=A$ we define 
\be 
\bm{A}=A_i\theta^i \quad \text{and} \quad \widetilde{\bm{A}}=A_i\chi^i~.
\ee
Then, in the same spirit as above, the corresponding Galileon is given as 
\bea
S_{2n}[A] =\frac 1 {(D-3n+1)!}\int\dd^{D}x \int \dd^{D} \theta \, \dd^{D} \chi \, \bm{\eta}^{D-3n+1} \,
\bm{\dd}\bm{A} \, \widetilde{\bm{\dd}}\widetilde{\bm{A}} \, 
(\bm{\dd}\widetilde{\bm{\dd}}\widetilde{\bm{A}} )^{n-1} \,
(\widetilde{\bm{\dd}}\bm{\dd} \bm{A} )^{n-1}~. 
\label{gg1form}
\eea
The proof of the statement follows the same steps as in \eqref{proofscalar}. Moreover, it is known from \cite{pformgalileons} that 
this action is trivial for $n>1$. Indeed this is now easily seen, since
in the graded formalism this conclusion is simply due to the fact that $\bm{A}$ and $\widetilde{\bm{A}}$ have degree 1 and thus 
(see Appendix A)
\be \label{dda}
(\bm{\dd}\widetilde{\bm{\dd}}\bm{A})^2=(\bm{\dd}\widetilde{\bm{\dd}}\widetilde{\bm{A}})^2=0~.
\ee
However, for $n=1$ this is certainly not true. This is expected since we know at least one non-trivial theory for an Abelian 
1-form with second order field equations, namely 
electromagnetism. Its action is thus included in \eqref{gg1form}; standard electrodynamics in e.g.~$D=4$ may be written in the graded formalism as 
\be 
S_{\text{Maxwell}}[A]=-\frac 12\int \dd^4x\int \dd^{4}\theta \, \dd^{4}\chi \,  \bm{\eta}^{2}\, \bm{\dd}\bm{A}\, 
\widetilde{\bm{\dd}}\widetilde{\bm{A}}~,
\ee
which amounts to the usual Maxwellian kinetic term.

The above action for vector fields is invariant under the symmetry
 \begin{align}
    \delta {\bf A} = {\bf d} \lambda + b_{ij} x^i \theta^j \,.
 \end{align}
The first term is just a usual gauge transformation, while the second part does not leave the Maxwell field strength $\bf d A$ invariant - instead, this transforms under the antisymmetric part of the constant parameter $b$. However, since the above action leads to purely second order field equations, these will be invariant under such transformations. Therefore the first term of the transformation above, with local parameter $\lambda$, is the generalization of the shift symmetry of scalars to forms. The second part, with global parameter $b$, is the analogon of the Galileon symmetry. Theories such as Born-Infeld, containing arbitrary functions of the Maxwell kinetic term, will only be invariant under the first part of this symmetry. 

In the general case of a $p$-form $\omega$ 
we use the expressions \eqref{omegatheta} and 
\eqref{omegachi} for $\bm{\o}$ and $\widetilde{\bm{\o}}$ respectively. 
Then the $p$-form Galileon is given by the same action \eqref{gg1form} with the substitution $3n \to (p+2)n$. 
Moreover, for $n=1$ one obtains the standard kinetic term $ F_{i_1\dots i_{p+1}}F^{i_1\dots i_{p+1}}$ for an Abelian $p$-form 
with $(p+1)$-form field strength.
For odd $p$ and $n>1$ this action is a total derivative for the same reason as for $p=1$, cf. \cite{pformgalileons}, 
in our case due to the graded identities
\be \label{oddzero}
(\bm{\dd}\widetilde{\bm{\dd}}\bm{\omega^{(\text{odd})}})^2=
(\bm{\dd}\widetilde{\bm{\dd}}\widetilde{\bm{\omega}}^{(\text{odd})})^2=0~,
\ee
which are proven similarly to the 1-form case. Indeed they are just the consequence of the fact that an odd differential form 
has odd grading, while the operator $\bm{\dd}\widetilde{\bm{\dd}}$ has grading 2. Thus $\bm{\dd}\widetilde{\bm{\dd}}\bm{\omega}$ 
has odd grading and its square vanishes.

Note that the general $p$-form action always has an even number of second derivative terms (multiplying the leading single derivative terms), while this is not the case for the scalar action \eqref{ggscalar}. Indeed, the latter, which can be seen as the $p=0$ case, allows for odd numbers as well. These additional possibilities arise since the two forms $\bf \omega$ and ${\bf \tilde \omega}$ coincide in this case.

A  similar action may be written for a collection of an arbitrary number of Abelian differential form species in flat space-time{\footnote{
Early attempts to generalize the scalar Galileons were made in \cite{SYZhou2}. 
}}.
Let us assume that we have a tower of $n$ gauge fields of degree 
$p_1,\dots, p_{n}$ respectively; these degrees may be all different or some of them may be equal as well.
Then the general Galileon action with strictly second order field equations can be written as 
\bea 
S[\o_1,\dots,\o_n]&=&\frac 1{(D-k)!}\int\dd^Dx\int\dd^D\theta\, \dd^D\chi \,  \bm{\eta}^{D-k} 
\bm{\dd}\bm{\o}_1^{(p_1)}\, \widetilde{\bm{\dd}}\widetilde{\bm{\o}}_2^{(p_2)}\times \nn\\
&& \qquad \qquad \qquad \qquad  \times \,\prod_{r=3}^{n_1} \bm{\dd}\widetilde{\bm{\dd}} 
\bm{\o}_r^{(p_r)}\prod_{s=n_1+1}^{n} \bm{\dd}\widetilde{\bm{\dd}}\widetilde{\bm{\o}}_s^{(p_s)}~,
\label{pcoupled}\eea
for some $n_1\le n$.
Provided that there are no more than one appearances of second derivatives of the same odd-graded field, this action is not a total derivative. Moreover, in order to have the required number of $\theta$s and $\chi$s, one has to impose the conditions
\be \label{matching}
p_1+\sum_{r=3}^{n_1}p_r = p_2 + \sum_{r=n_1+1}^np_r = k - n +1 ~.
\ee  
For instance, when all species are $p$-forms of equal degree, the condition \eqref{matching} is satisfied if and only if 
$n_1=\tfrac{1}{2}(n+2)$, which means that there is an even number of terms coming in pairs (provided $p \neq 0$). This is in accord to our previous considerations. 

The leading factors of $\bm{\eta}$ in \eqref{pcoupled} are related to Hodge duality. Indeed, considering $p$-forms $A$ and $B$, we find
\bea
\frac1{(D-p)!}\bm{\eta}^{D-p} \widetilde{\bm A}^{(p)} {\bm B}^{(p)}& =& 
\frac1{(D-p)!}\frac1{(p!)^2}\, \varepsilon_{j_1\ldots j_{D-p}}{}^{i_1\ldots i_p}  A_{i_1\ldots i_p} 
 B_{j_{D-p+1}\ldots j_D} \theta^{j_1}\ldots\theta^{j_D} \chi^1\ldots \chi^D\nonumber\\
&= &(\star \bm A \wedge \bm B)^{(D)}  \chi^1 \ldots \chi^D~.
\label{Hodge1}
\eea

To close this discussion of coupled $p$-form Galileons, we note that in \cite{pformgalileons} the following cases were analysed 
explicitly:
\begin{itemize}
 \item $ p_1=p_2=1, p_3=0, n= n_1=3$. This is a simple case of a coupled system containing one scalar $\pi$ and one 1-form $A$. 
 The action in $D\ge k=3$ dimensions is given by 
 \be 
 \frac 1{(D-3)!}\int\dd^Dx\int\dd^D\theta \, \dd^D\chi \, \bm{\eta}^{D-3}\, \bm{\dd}\bm{A}\,\widetilde{\bm{\dd}}\widetilde{\bm{A}}\,\bm{\dd}\widetilde{\bm{\dd}}\pi~.
 \ee
 \item $ p_1=p_2=0, p_3=p_4=1, n=4, n_1=3$. This is yet another coupled system of a single scalar and a single $1$-form 
 in $D\ge k=4$, with action
  \be 
 \frac 1{(D-4)!}\int\dd^Dx\int\dd^D\theta \, \dd^D\chi \, \bm{\eta}^{D-4}\, \bm{\dd}\pi\,
 \widetilde{\bm{\dd}}\pi\,\bm{\dd}\widetilde{\bm{\dd}}\widetilde{\bm{A}}
 \, \widetilde{\bm{\dd}}\bm{\dd}\bm{A}~.
 \ee
\end{itemize}

Finally, as an illustration and further generalization, let us consider two actions for a single 3-form $A = \frac16 A_{ijk} \dd x^i \wedge \dd x^j \wedge \dd x^k$ in nine dimensions that differ in the way how $A$ is contracted by $\theta$'s and $\chi$'s. Using only $\bm A = \frac16 A_{ijk} \theta^i\theta^j\theta^k$
and its tilde-dual $\widetilde{\bm A} = \frac16 A_{ijk} \chi^i\chi^j\chi^k$ leads to the trivial action
\bea 
S[A]&=&\int\dd^9x\int\dd^9\theta\, \dd^9\chi \, 
  \bm\dd \widetilde{\bm\dd}\widetilde{\bm A} \, \widetilde{\bm\dd}\bm\dd\bm A \,
 \bm\dd\widetilde{\bm A} \,
 \widetilde{\bm\dd}\bm A~.
\eea
Instead, considering also the other contraction $\bm A' =\frac12  A_{ijk} \theta^i\theta^j\chi^k$ and its tilde-dual $\widetilde{\bm A'} =\frac12  A_{ijk} \chi^i\chi^j\theta^k$, both of which employ the same anti-symmetric three-form components, yields the non-trivial action
\bea 
S[A]&=&\int\dd^9x\int\dd^9\theta\, \dd^9\chi \,  
  \bm\dd \widetilde{\bm\dd}\widetilde{\bm A} \, \widetilde{\bm\dd}\bm\dd\bm A \,
 \bm\dd\widetilde{\bm A'} \,
 \widetilde{\bm\dd}\bm A' \,,
\eea
that has been previously considered in \cite{pformgalileons2}. Note that this generalization breaks the gauge invariance discussed above.

\section{Galileons for mixed-symmetry tensor fields}
\label{msgal}

\subsection{Preliminaries on ${(p,q)}$ mixed-symmetry tensors}

Mixed-symmetry tensor fields were   studied by Curtright as generalizations of gauge fields \cite{curtright}. They are tensors of total degree 
$m\in\Z_+$ with more than one 
set of separately antisymmetric indices, corresponding to Young diagrams for different integer partitions of $m$. 
Although these partitions may certainly contain more than two summands, for the purposes of this paper we focus on the 
simplest partitions involving just two summands
\be 
m=p+q~,\quad p,q\in \Z_{+}~.
\ee
In this case we speak of a $(p,q)$ mixed-symmetry tensor.\footnote{Let us stress that
here we work with covariant tensors without any contravariant parts.}

Mixed-symmetry tensors may be understood as multi-forms, namely tensor products of differential
forms of generically different 
degree, which is the approach considered in \cite{demedeiroshull1}. In this sense, here we focus on bi-forms. It turns out that this case is the most general one we can consider for our purposes. 
Recall that we are guided by the requirement of second order field equations. Had we allowed more summands in the partition of 
$m$, this would have violated this fundamental assumption, as it will become obvious below.

Due to its nature, such a tensor can be 
represented in a local coordinate basis $(x^i)$ on space-time $M$ as 
\be 
T^{(p,q)}=\sfrac 1{p!q!}T_{[i_1\dots i_p][j_1\dots j_q]}\dd x^{i_1}\otimes\dots\otimes\dd x^{i_p}\otimes\dd x^{j_1}\otimes\dots\otimes\dd x^{j_q}~.
\ee
Defining the usual antisymmetric product of differential 1-forms,
\be
\omega\w\omega'=\omega\otimes\omega'-\omega'\otimes\omega~, \quad \omega,\omega'\in\Omega^1(M)~,
\ee
a $(p,q)$ mixed-symmetry tensor may be equivalently represented as 
\be 
T^{(p,q)}=\sfrac 1{p!q!}T_{i_1\dots i_pj_1\dots j_q}\dd x^{i_1}\w\dots\w\dd x^{i_p}\otimes \dd x^{j_1}\w\dots\w\dd x^{j_q}~.
\ee
The components of GL($D$)-\emph{irreducible} mixed symmetry tensors satisfy in addition the following two conditions:
\bea 
T_{[i_1\dots i_pj_1]\dots j_q}&=&0~,\label{vanishingextraantisym}
\\
 T_{[i_1\dots i_p][j_1\dots j_q]}&=&T_{[j_1\dots j_q][i_1\dots i_p]}~, ~ \text{for} ~p=q~.\label{symmetry}
\eea
Then \eqref{vanishingextraantisym} means that there are no additional antisymmetric properties for the tensors, while \eqref{symmetry} 
states that for $p=q$ there is an additional symmetry under interchanging both sets of mixed-symmetry fields. Most of what follows is valid also for reducible tensors, i.e. without having to impose these two conditions. We will state explicitly whenever \eqref{vanishingextraantisym} and \eqref{symmetry} are assumed.

The cases $m=0$ and $m=1$ are degenerate, containing only scalar fields and 1-forms respectively. The first non-trivial case 
arises when $m=2$, whence the two possibilities are $(p,q)=(2,0) $ and $(p,q)=(1,1)$. (Note that there is also an overall symmetry under the interchange of all $\chi$'s and $\theta$'s. Therefore, there is no need to study $(p,q) = (0,2)$.) The former case corresponds to a 2-form 
and it is already studied in \cite{pformgalileons}. However the latter case is not covered by the analysis of \cite{pformgalileons}. If we focus on irreducible $(1,1)$ tensor fields, this corresponds to a spin-2 field, essentially a massless graviton. Similarly, for $m=3$ one obtains a 3-form as well as a 
$(2,1)$ mixed-symmetry tensor with components $T_{[i_1i_2]j_1}$ (in addition to a fully symmetric $(1,1,1)$ case, which corresponds to a partition of $m$ into three 
summands and thus is not covered by the framework that we consider here.) For $m=4$, apart from the 4-form there are    additional mixed-symmetry tensors of 
type $(3,1)$ and $(2,2)$.

In the spirit of the graded formalism introduced in the previous section, 
 $(p,q)$ mixed-symmetry (reducible or irreducible) tensors acquire the following simple representations in terms of  the degree-1 Grassmannian variables:
\bea 
\bm{\o}^{(p,q)}&=&\sfrac 1{p!q!}\o_{i_1\dots i_pj_1\dots j_q}\theta^{i_1}\dots\theta^{i_p}\chi^{j_1}\dots \chi^{j_q}~,\label{tpq}
\\
\widetilde{\bm{\o}}^{(q,p)}&=&\sfrac 1{p!q!}\o_{i_1\dots i_pj_1\dots j_q}\chi^{i_1}\dots\chi^{i_p}\theta^{j_1}\dots \theta^{j_q}~.\label{ttpq}
\eea
The first entry in the label denotes the number of $\theta$'s, the second denotes the number of $\chi$'s. The tilde-duality operation ``$\widetilde{\phantom \o}$'' replaces all $\theta$s with $\chi$s and vice versa and thus maps the label $(p,q)$ to $(q,p)$.

Important for our purposes are exterior derivatives on $(p,q)$ mixed-symmetry tensors. 
In an obvious generalization of $p$-forms, there are now in fact two independent exterior derivatives
\be 
\dd : \Omega^{(p,q)}(M)\to \Omega^{(p+1,q)}(M)~ \quad \text{and} \quad \widetilde\dd :
 \Omega^{(p,q)}(M)\to \Omega^{(p,q+1)}(M)~,
\ee
acting accordingly as 
\bea 
\dd \o&=&\sfrac 1{p!q!}\partial_{i_0}\o_{i_1\dots i_pj_1\dots j_q}\dd x^{i_0}\w\dd x^{i_1}\w\dots\w\dd x^{i_p}\otimes\dd x^{j_1}\w\dots\w\dd x^{j_q}~,\\
\widetilde\dd \o&=&\sfrac 1{p!q!}\partial_{j_0}\o_{i_1\dots i_pj_1\dots j_q}\dd x^{i_1}\w\dots\w\dd x^{i_p}\otimes\dd x^{j_0}\w\dd x^{j_1}\w\dots\w\dd x^{j_q}~.
\eea
The above derivatives satisfy the following useful properties:
\be 
\dd^2=\widetilde\dd^2=0 \quad \text{and} \quad \dd\widetilde\dd=\widetilde\dd\dd~,
\ee 
namely they are nilpotent of degree 2 and they commute. As before, one could use different conventions such that the two derivatives
anti-commute. These derivations satisfy exactly the same properties as their graded counterparts $\bm{\dd}$ and 
$\widetilde{\bm{\dd}}$.

\subsection{The mixed-symmetry Galileon for a single species}

Our purpose here is to write a general higher-derivative action for a $(p,q)$ mixed-symmetry tensor field such that it leads 
to exactly second order field equations. 
Thus we consider such a $(p,q)$ mixed-symmetry tensor field $\o^{(p,q)}$ and its graded counterparts given in 
Eqs. \eqref{tpq} and \eqref{ttpq}. Then 
 the action is
\bea 
 S_{2n}[\o]&=&\frac 1{(D-k)!}\int\dd^D x\int\dd^D\theta\, \dd^D\chi\,\bm{\eta}^{D-k} \bm{\dd}\bm{\o}\, 
\widetilde{\bm{\dd}}\widetilde{\bm{\o}}\, (\bm{\dd}\widetilde{\bm{\dd}}\bm{\o})^{n-1}\,
(\bm{\dd}\widetilde{\bm{\dd}}\widetilde{\bm{\o}})^{n-1}~,
\eea
with $k=(p+q+2)n-1$. We observe that the action is formally the same as the one for $p$-forms, 
but now holds more generally for mixed-symmetry tensors. Variation with respect to $\bm{\o}$ directly 
shows that the field equations are strictly second order in derivatives. 
 
The above action is gauge invariant (up to boundary terms)
provided that $\bm{\dd}\widetilde{\bm{\dd}}(\d \bm{\o})=0$. For $p,q \geq 1$ this implies (in a contractible patch)
\be 
\fbox{$\displaystyle
\d\bm{\o}^{(p,q)}=\bm{\dd}\bm{\lambda}^{(p-1,q)}+\widetilde{\bm{\dd}}{\bm{\l'}}^{(p,q-1)} +  b_{i_0 i_1 \ldots i_{p+q}} x^{i_0} \theta^{i_1} \cdots \theta^{i_p} \chi^{i_{p+1}} \cdots \chi^{i_{p+q}}
~,$} \label{gen-Gal}
\ee
with a  \emph{constant totally antisymmetric} $p+q+1$ tensor $b_{i_0 i_1 \ldots i_{p+q}}$. The proof is done by induction over either $p$ or $q$ and repeatedly uses the ordinary Poincar\'e lemma. If either $p=0$ or $q=0$ the respective gauge parameter $\bm \lambda$ or $\bm {\lambda'}$ should be omitted. For $p=q=0$  (i.e.\ scalar Galileons) the gauge transformation terms $\bm{\dd}\bm{\lambda}+\widetilde{\bm{\dd}}{\bm{\l'}}$ should be replaced by a single constant~$c$. This can be summarized as
\be 
\d\bm{\o}^{(p,q)}=
\left\{
\begin{array}{lr}
\bm{\dd}\bm{\lambda}^{(p-1,q)}+\widetilde{\bm{\dd}}{\bm{\l'}}^{(p,q-1)} +  b_{i_0 i_1 \ldots i_{p+q}} x^{i_0} \theta^{i_1} \cdots \theta^{i_p} \chi^{i_{p+1}} \cdots \chi^{i_{p+q}}  & (p,q> 0)\\
\bm{\dd}\bm{\lambda}^{(p-1,0)} +  b_{i_0 i_1 \ldots i_{p}} x^{i_0} \theta^{i_1} \cdots \theta^{i_p}   & (p>0, q= 0)\\
\widetilde{\bm{\dd}}{\bm{\l'}}^{(0,q-1)} +  b_{i_0 i_1 \ldots i_{q}} x^{i_0}  \chi^{i_{1}} \cdots \chi^{i_{q}}  & (p= 0,q>0)\\
c +  b_{i} x^{i}   & (p=q= 0)
\end{array}
\right.
\label{genmult-Gal}
\ee
where $b$, $c$ are constant and $b$ is fully antisymmetric in all of its indices.  

While for $p,q \geq 1$  the first two terms are gauge transformations with local parameters, we would like to argue that this is the natural generalization of the Galileon transformation of scalars to the case of bi-forms. In particular, the first derivative of the tensor is not invariant under such a transformation, since ${\bf d \tilde d} \neq 0$. Only the second derivatives of the tensor are invariant, similar to the Galileon case. This again emphasizes the close connection to the two sets of Grassmann variables and the second order nature of the field equations. Moreover, it stresses the striking similarity between Galileon scalars and bi-forms with the above gauge invariance.

Let us now consider GL($D$)-irreducible tensors $\bm \omega$ and restrict the above transformations to those that do not leave this category. The last term in \eqref{gen-Gal} does not respect \eqref{vanishingextraantisym} for  $p,q \geq 1$, so we are left with the first two gauge transformation-like terms.
An important special case arises when $p=q$, in which irreducibility trivially implies $\bm \o = \tT$. Similar to the scalar case, the structure of the action is enhanced in this case, allowing for more 
possibilities. Specifically, it acquires the following form:
\be
\label{addaction}
S_{n+1}[\o^{(p,p)}]=\frac 1{(D-k)!}\int\dd^D x\int\dd^D\theta\, \dd^D\chi\,\bm{\eta}^{D-k} \bm{\dd}\bm{\o}\, 
\widetilde{\bm{\dd}}{\bm{\o}}\, (\bm{\dd}\widetilde{\bm{\dd}}\bm{\o})^{n-1}
~,
\ee
with $k=(p+1)n+p$. Let us highlight here the following important difference. 
In the general case, for $n=1$ one obtains an action with 2 appearances of the field and for $n=2$ one with 4 appearances of the field. 
A 3-field case, or any odd number for that matter, is not possible. On the contrary, the special $p=q$ case allows for odd field appearances.

Let us now turn to some examples. The first example of a mixed-symmetry tensor field corresponds to $m=2$ and the partition $(p,q)=(1,1)$. The relation of this case with gravity will be discussed in the next subsection. 
Let us also mention an example where $p\ne q$. We could start with the 
simplest $p\ne q$ case of $(p,q)=(2,1)$. However, the action is then a total derivative, unless $n=1$. The proof is similar to 
the differential form case and generalizes to any mixed-symmetry tensor with odd $p+q$. In particular, once more it holds that 
\be \label{oddzero2}
(\bm{\dd}\widetilde{\bm{\dd}}\bm{\o}^{(p,q)})^2|_{p+q=\text{odd}}=
(\bm{\dd}\widetilde{\bm{\dd}}\widetilde{\bm{\o}}^{(p,q)})^2|_{p+q=\text{odd}}=0~.
\ee
Thus the first non-trivial single-field case with $p\ne q$ arises for $(3,1)$. One can then write the action in 11 dimensions 
for $n=2$, first in the compact graded formalism and then in index notation,
\bea 
S_{4}[\o^{(3,1)}]&=&\int\dd^{11} x\int\dd^{11}\theta\, \dd^{11}\chi\,\ \bm{\dd}\bm{\o}^{(3,1)}\widetilde{\bm{\dd}}
\widetilde{\bm{\o}}^{(1,3)}\bm{\dd}\widetilde{\bm{\dd}}\bm{\o}^{(3,1)}
\bm{\dd}\widetilde{\bm{\dd}}\widetilde{\bm{\o}}^{(1,3)}\nn\\
&=&
\int\dd^{11}x ~\vare^{i_1\dots i_{11}}\vare^{j_1\dots j_{11}}
\partial_{i_1}\o_{i_2i_3 i_4j_1}
\partial_{j_2}\o_{j_3 j_4 j_5 i_5} 
\partial_{i_6}\partial_{j_6}\o_{j_7 j_8 j_9 i_7}
\partial_{i_8}
\partial_{j_{10}}\o_{i_9 i_{10} i_{11} j_{11}}~.\nn
\eea
The second line indicates the non-trivial distribution of indices in the two antisymmetric tensors. Then the first line clarifies the 
elegance of the graded formulation, where one does not have to worry about the position of the indices.

\subsection{Relation to Lovelock gravity}

Now let us turn to the $(1,1)$ case which is special due to its connection with gravity. We will denote the corresponding gauge field as $h:=\o^{(1,1)}$ with 
\be 
h=h_{ij}\dd x^i\otimes \dd x^j~, \quad \bm{h}=h_{ij}\theta^i\chi^j~.
\ee
This is nothing but a massless spin-2 field, and as such it is the simplest of the special cases with $p=q$. 
It is therefore expected that in this case one can relate the Galileon-type action \eqref{addaction} to linearised Lovelock gravity, which we will shortly recap. 

The most general action for a metric theory of gravity in $D$ dimensions with second order conserved equations of motion can be written as a sum \cite{lovelock}
\be 
S_{\text{Lovelock}} = \int d^D x \, \sum_{n=0}^{\lfloor \sfrac {D-1}2\rfloor} \alpha_n \mathcal L_n
\ee
over the dimensionally extended Euler densities
\be \label{LoveGal1}
\mathcal L_n = \frac{\sqrt{-g}}{2^n} \delta^{k_1l_1 \ldots k_n l_n}_{i_1j_1\ldots i_nj_n} \prod_{r=1}^n R^{i_r j_r}{}_{k_rl_r} \, ,
\ee
where $\alpha_n$ are constants (and in particular the cosmological constant for $n=0$).
These Lovelock invariants can be neatly re-expressed in the graded formalism as
 products of powers of  $\bm{g} =g_{ij}\theta^i \chi^j$ and $\bm{Riem} =
\theta^i \theta^j \chi^k \chi^l R_{ijkl}$:
\be \label{LoveGal}
\mathcal L_n  = \frac{\sqrt{-g}}{(D-2n)!} \int d^D\theta d^D\chi \,\bm{g}^{D-2n} \left(\sfrac 12 \bm{Riem}\right)^n \, .
\ee
Indeed, integrating over $\theta$ and $\chi$ and using the $D-2n$ copies of the metric to lower indices, we obtain from  (\ref{LoveGal}):
\be
\mathcal L_n  = \frac{\sqrt{-g}}{(D-2n)!} \epsilon^{k_1l_1 \ldots k_nl_n s_{2n+1}\ldots s_D} \epsilon^{i_1j_1\ldots i_nj_n}{}_{s_{2n+1}\ldots s_D}
\prod_{r=1}^n \sfrac 12 R_{i_r j_rk_rl_r} \,,
\ee
which is equal to (\ref{LoveGal1}).

Turning to the linearised version of these fully non-linear gravitational theories, we will think of $h$ 
as a symmetric tensor fluctuation to the Minkowski metric such that $h_{ij}(x)\ll 1$.  The linearised Einstein-Hilbert action in four dimensions becomes
\bea
S_{\text{LEH}}[h] 
=
-\frac{1}{2} \int\dd^4 x \; h^{ij}
\left( {R}_{ij} - \sfrac{1}{2} \eta_{ij} {R} \right)
~,\label{lehaction}
\eea
where
${R}_{ij} = {R}^{k}_{\phantom{k}ikj}$ is the linearised Ricci tensor and ${R} = \eta^{ij}{R}_{ij}$ is the linearised Ricci scalar. Both follow from the linearised Riemann curvature   expressed in terms of linearised metric:
\bea
{R}_{i_4 i_3 j_3 j_4}
= \frac{1}{2} \left(\partial_{i_3} \partial_{j_3} h_{i_4 j_4}
+ \partial_{i_4} \partial_{j_4} h_{i_3 j_3}
- \partial_{i_3} \partial_{j_4} h_{i_4 j_3}
- \partial_{i_4}\partial_{j_3} h_{i_3 j_4} \right)~,
\eea
or, in graded notation, $\bm{R} = R_{i_1i_2j_1j_2} \theta^{i_1}\theta^{i_2}\chi^{j_1}\chi^{j_2}  = -2 \bm{\dd}\widetilde{\bm{\dd}}\bm{h} $.
Indeed, one can express this as 
\bea 
S_{\text{LEH}}[h]= { -\frac{1}{4}} \int\dd^4 x\int\dd^4\theta\, \dd^4\chi\,  \bm{\eta} \, \bm h
 \,\bm{\dd}\widetilde{\bm{\dd}}\bm{h}~, 
\eea
in our graded formalism, coinciding with the lowest order Galileons for a $(1,1)$ field. This is in fact the massless Fierz-Pauli action.

Moving on to cubic interactions, the lowest number of dimensions that we encounter a non-trivial action 
with more than two fields is five. 
In five dimensions there exists
a non-trivial Lovelock invariant beyond the Einstein-Hilbert term, the Gauss-Bonnet invariant. 
Again the linearised Lovelock invariant, in this case the Gauss-Bonnet term, coincides with the Galileon invariant for $h$, in this case at cubic order:
\bea\label{gbflat}
S_{\text{LGB}}[h]&=& -\frac{1}{4} \int\dd^5 x\int\dd^5\theta\, \dd^5\chi \, \bm h
 \,(\bm{\dd}\widetilde{\bm{\dd}}\bm{h})^2 
 \nn\\
 &=&
 -\frac{1}{16}  \int\dd^5 x \, 
 \varepsilon^{i_1 i_2 i_3 i_4 i_5} \varepsilon^{j_1 j_2 j_3 j_4 j_5} \,
 h_{i_1 j_1} \, {R}_{i_2 i_3 j_2 j_3} \, {R}_{i_4 i_5 j_4 j_5} \nn\\
 &=&
\frac 12 \int\dd^5 x \, h^{ij}\left(H_{ij}-\sfrac 12\eta_{ij}{\cal L}_{\text{GB}} \right)~,
\eea
where 
\bea 
{\cal L}_{\text{GB}}&=&{R}^2 + {R}_{ijkl} {R}^{ijkl}
- 4 {R}_{ij} {R}^{ij}~,
\\
H_{ij}&=&2( R_{ij} R - 2 R_{ikjl} R^{kl} - 2R_{ik}R^k_{\phantom{k}j} + R_i^{\phantom{i}mkl} R_{jmkl} )~,
\eea
both involving linearised curvature tensors in the present context. Note that the tensor $H_{ij}$ appears in the field equations 
derived from the Gauss-Bonnet term \cite{Deruelle:2003ck}. Its trace yields the Gauss-Bonnet term, in the same way that the Ricci tensor yields the curvature scalar.  

In general, for any space-time dimension
$D \ge 2n+1$,
\bea
-\frac{1}{4}  S_{n+1}[h]=S_n^{\text{LL}}[h]~,
\eea 
where $S_{n+1}$ on the left is the action (\ref{addaction}) for $p=1$ and $S_n^{\text{LL}}$ is the linearised $n$-th order Lovelock action\footnote{
A connection between the Lovelock invariants and covariant Galileons 
has been discussed in \cite{Acoleyen} in a Kaluza-Klein compactification setup.}:
\bea
S_n^{\text{LL}}[h]
&=& 
-\frac14   \frac{1}{(D-2n-1)!}  \left(-\frac{1}{2}\right)^n \int\dd^D x 
\int\dd^D\theta\, \dd^D\chi \, \bm{\eta}^{D-2n-1} \, \bm h \,
 \bm{ R}^n  \nn \\
&=&
-\frac14   \frac{1}{(D-2n-1)!} 
\int\dd^D x\int\dd^D\theta\, \dd^D\chi \, \bm{\eta}^{D-2n-1} \, \bm h
 \, (\bm{\dd}\widetilde{\bm{\dd}}\bm{h} )^n
 ~.
\eea
This is the lowest order expression in $h$ with a 
non-trivial action. (The version with $D-2n$ powers of $\eta$ and no 
lone $h$ is a total derivative.)

We conclude that the Galileon-type action \eqref{addaction} yields an elegant formula for the linearised version of Lovelock terms in any dimension. Moreover, the gauge transformations of tensor Galileons in this case correspond to the linearised version of diffeomorphisms. Thus gravity can be seen as a non-linearised completion of the first non-trivial tensor Galileons with $(p,q)=(1,1)$.

\subsection{Multiple species and generalized mixed-symmetry Galileons}

It is now fairly obvious what is the Galileon action with strictly second order field equations in the case 
of an arbitrary number, say $n$, of mixed-symmetry tensor fields $\o_k^{(p_k,q_k)}$ (including the degenerate cases of differential forms and scalar 
fields as special ones) 
of any degree:
\bea 
S[\o_0,\dots,\o_n]&=&\frac 1{(D-k)!}\int\dd^Dx\int\dd^D\theta\, \dd^D\chi \,  \bm{\eta}^{D-k} 
  \bm{\o}_0^{(p_0,q_0)}
\prod_{j=1}^{n} \bm{\dd}\widetilde{\bm{\dd}} 
\bm{\o}_j^{(p_j,q_j)} ~.
\label{pqcoupled}\eea
As before, the criterion based on the identities 
\eqref{oddzero2} controls when the action is non-trivial. Moreover, the action is non-vanishing only for the correct number of $\theta$s 
and $\chi$s, which in the present case boils down to the conditions
\bea \label{conditiongeneral}
\sum_{k=0}^{n} p_k = \sum_{k=0}^n q_k= k-n ~.
\eea
The tensor fields $\o_k$ in \eqref{pqcoupled} do not all need to be different, some or all may coincide, or could be tilde-duals (see \eqref{tpq}, \eqref{ttpq}) of one another, or they could be related by Hodge duality, etc..
The action \eqref{pqcoupled} is invariant (up to boundary terms) under the Galilean-style transformations \eqref{genmult-Gal} of any of its fields $\o_k^{(p_k,q_k)}$. 

In \eqref{pcoupled} the leading factors of $\bm \eta$ were related to the usual Hodge duality of differential forms, see equation \eqref{Hodge1}. In the present notation this reads $(\star \bm \omega)^{(D-p,D)} = \frac1{(D-p)!}\bm \eta^{D-p} {\widetilde {\bm \omega}}^{(0,p)}$, where a factor $\chi^1 \ldots \chi^D$ is included in  $(\star\bm \omega)^{(D-p,D)}$. The appropriate generalization of Hodge duality to a mixed-symmetry $(p,q)$-tensor field with $p+q \leq D$ is
\be
(\star {\bm\omega})^{(D-p,D-q)} = \frac1{(D-p-q)!}{\bm \eta}^{D-p-q} {\widetilde{\bm \omega}}^{(q,p)}
\ee
and is in general \emph{not} equal to naive Hodge dualization in each of the two tensor factors. 
In fact, this prescription works not only for individual tensor fields but also for composite expressions involving several fields of overall degree $(p,q)$, i.e.\ including interacting fields.  

Let us now examine a couple of non-trivial examples. Starting with $n=3$, we consider three mixed-symmetry tensor fields 
of different degree, $(2,1), (3,1)$ and $(3,2)$ respectively. This guarantees that 
\eqref{conditiongeneral} is satisfied. Moreover, the minimal amount of dimensions for a coupling of these fields is 
8. To avoid confusion we use the notation $\o_1^{(2,1)}=A$, $\o_2^{(3,1)}=B$ and $\o_3^{(3,2)}=C$. The corresponding Galileon is then simply given as 
\be
\label{ABC}
S[A,B,C]=\int\dd^8x\int\dd^8\theta\, \dd^8\chi \, \bm{\dd}\bm{A}\, \widetilde{\bm{\dd}}\widetilde{\bm{B}}\,
 \bm{\dd}\widetilde{\bm{\dd}} 
\bm{C}~,
\ee
which in the indexed formalism translates into
\be
S[A,B,C]=\int\dd^8x \,\vare^{i_1\dots i_8}\vare^{j_1\dots j_8}\, \partial_{i_1}A_{i_2i_3j_1}\, \partial_{j_2}B_{j_3j_4j_5i_4}
\,
 \partial_{i_5}\partial_{j_6}C_{i_6i_7i_8j_7j_8}~,
\ee
which shows once more the non-trivial distribution of indices in the antisymmetric tensors.
As a second example, let us consider another case with $n=3$ and couple a scalar field $\pi$  with a mixed-symmetry 
tensor $N$ with degree $(7,1)$. By counting, the relevant action is one in 10 dimensions and it reads as 
\be 
\label{piN}
S[\pi,N]=\int\dd^{10}x\int\dd^{10}\theta\, \dd^{10}\chi \, \bm{\dd}\pi\, \widetilde{\bm{\dd}}\widetilde{{\bm{N}}}\,
 \bm{\dd}\widetilde{\bm{\dd}} 
\bm{N}~.
\ee
The indexed version of this action is
\be
S[\pi,N]=\int\dd^{10}x \, \vare^{i_1\dots i_{10}}\vare^{j_1\dots j_{10}} \partial_{i_1}\pi\, \partial_{j_1}
N_{i_2i_3i_4i_5i_6i_7i_8j_2}\, \partial_{i_{9}}\partial_{j_3}N_{j_4j_5j_6j_7j_8j_{9}j_{10}i_{10}}~.
\ee
An interesting feature of a mixed-symmetry tensor with $(7,1)$ degree is that in ten dimensions it is dual to the graviton, at 
least at the linearised level.

In the present formalism, it is now  simple to generalize the above actions in the case when the 
field equations are required to be \emph{up to} second order in derivatives. In particular, suppose we have a tower of $n$
mixed-symmetry tensor fields, as above. Requiring field equation with second or less derivatives and an action which is 
polynomial in all fields, we write
\bea 
S[\o_0,\dots,\o_n]&=&\frac 1{(D-k)!}\int\dd^Dx\int\dd^D\theta\, \dd^D\chi \,  \bm{\eta}^{D-k} 
\prod_{i}   \bm{\o}_i^{(p_i,q_i)} \times \nonumber \\
&&
 \times \prod_{j} \bm{\dd} \bm{\o}_j^{(p_j,q_j)}
\prod_{k} \widetilde{\bm{\dd}}\bm{\o}_k^{(p_k,q_k)} 
\prod_{l} \bm{\dd}\widetilde{\bm{\dd}} 
\bm{\o}_l^{(p_l,q_l)}~.
\label{pqcoupled2}
\eea
In order for terms in the action to be non-zero, all odd variables $\theta^1$, \ldots, $\theta^D$, $\chi^1$, \ldots, $\chi^D$ need to appear exactly once.
This action will in general no longer be invariant under the Galilean transformations \eqref{genmult-Gal}.

Of course this is not the most general action in case non-polynomial scalar couplings are allowed. This 
can be invoked already from the analysis of \cite{ggalileons}, generalized here to the case of an arbitrary tower of 
mixed-symmetry tensors. In case some of the species in the theory are scalar fields, say $\pi$,
one  may just include in \eqref{pqcoupled2} 
an arbitrary function $f(\pi,X)$, where $X=(\partial\pi)^2$. In this latter case one ends up with undifferentiated fields as well, apart from once and twice differentiated ones. Similarly, one can envision Born-Infeld-type of theories for any tensor field, containing interaction terms that consist of arbitrary functions of the canonical kinetic term for these.

\section{Covariant tensor Galileons}

We would now like to explore the generalization of the actions we wrote in the previous section for 
single mixed-symmetry tensor fields in curved space-time. This has proven possible in the case 
of scalar and $p$-form Galileons \cite{covariantgalileons,pformgalileons}. Thus one could anticipate that it would work smoothly in the 
mixed-symmetry case too. However, there are a couple of reasons that render this hypothesis a bit too quick. First, in the very 
simplest case of the $(1,1)$ field $h$, we encounter a massless spin-2 field. Thus the task would be to couple it to an 
external graviton $g_{\m\nu}$, which is independent of $h$.{\footnote{To avoid confusion, in this section $h$ is \emph{not} 
the linear fluctuation of $g$ around a flat background.}} However, it is known that under certain rather general conditions, it is 
impossible to couple two massless gravitons non-trivially \cite{Boulanger:2000rq}. We shall comment on this below. Secondly, although scalars and $p$-forms are 
already covariant fields, in the sense that the exterior derivative acting on them is already a covariant one, this is no longer true for 
mixed-symmetry tensor fields. This has non-trivial consequences as will become clear in the ensuing.

\subsection{Preliminaries on covariant tensors}

The Grassmann variables $\theta^i$ and 
$\chi^i$ transform like differentials $\dd x^i$ and the exterior 
derivatives $\dd$ and $\widetilde \dd$ hence map scalar fields to  
$1$-tensors. For mixed tensors we however need to introduce covariant 
derivatives $\nabla_i$.
Their action on  the Grassmann variables is
\begin{equation}
\nabla_i \theta^k = - \theta^j \Gamma_{ij}^k
\quad \text{and} \quad
\nabla_i \chi^k = - \chi^j \Gamma_{ij}^k~.
\end{equation}
Note that $\G^i_{jk}$ are the Christoffel symbols for the external metric $g$. The same holds for all covariant derivatives that 
appear in this section. For simplicity we assume in the following that the torsion ${T}_{ij}^k = 
\Gamma_{ij}^k - \Gamma_{ji}^k$ is zero.    For the covariant graded exterior derivatives $\bm\nabla = 
\theta^i \nabla_i$ and $\widetilde{\bm \nabla} =  \chi^i \nabla_i$ we obtain
\begin{eqnarray}
\bm\nabla \theta^k &=& - \theta^i \theta^j \Gamma_{ij}^k = 0~,
\qquad\,\,\,\,\, \widetilde{\bm\nabla} \chi^k = - \chi^i\chi^j {\Gamma}_{ij}^k = 0~,
\\ \label{LCconn1}
\bm\nabla \chi^k &=& -\theta^i\chi^j\Gamma_{ij}^k =: - \bm\Gamma^k~, \quad
\widetilde{\bm\nabla} \theta^k = -\chi^i\theta^j\Gamma_{ij}^k
= - {\bm\Gamma}^k~.
\end{eqnarray}
Acting once more with either of the covariant derivatives gives terms 
involving the Riemann curvature tensor as can be seen from the relations
\begin{eqnarray}
\bm\nabla\bm\Gamma^j &=& \frac{1}{2} \chi^i\theta^k\theta^{\ell} 
R^{j}_{\phantom{j}ik\ell} 
=:  \bm R^j
\\
\widetilde{\bm\nabla} {\bm\Gamma}^j &=&
\frac{1}{2}\theta^i\chi^k\chi^{\ell} 
R^{j}_{\phantom{j}ik\ell} 
=: \widetilde{\bm R}^j ~.
\end{eqnarray}
The difference to the derivations formed with the partial derivative is evident. 
In the present case nilpotency is violated; the square and commutator of the 
covariant derivatives are
\begin{eqnarray}
{\bm\nabla}^2 = \sfrac 12 \theta^i\theta^j[\nabla_i,\nabla_j] =: \bm R 
\quad \text{and} \quad
 \widetilde{\bm\nabla}^2 = \sfrac 12 \chi^i\chi^j[\nabla_i,\nabla_j] =: \widetilde{\bm R}~,
\end{eqnarray}
\begin{equation}
[\bm\nabla, \widetilde{\bm\nabla}] = \chi^i\theta^j R_{ij} = :\bm R^{'} ~,
\end{equation}
where $\bm R$, $\widetilde{\bm R}$ and $\bm R^{'}$ are understood as curvature operators 
acting on tensors.
Moreover, recall that we have previously defined
\be
\bm{Riem} = R_{ijkl}\theta^i\theta^j\chi^k\chi^l~,\\
\ee
with the components of the Riemann tensor; 
the symmetries, e.g. antisymmetrization of the first two and the last two indices, are 
naturally implemented.

On $(p,0)$ tensors, namely standard differential $p$-forms, $\bm \nabla$ reduces to $\bm {\dd}$ and the same is true for 
$\widetilde {\bm \nabla}$ on $(0,q)$ tensors (assuming vanishing torsion). For all 
other tensors we do need in general non-trivial connection coefficients 
$\Gamma_{ij}^k$.

\subsection{Covariantization of the  mixed-symmetry Galileon}

As clarified above, the main difference in the case of mixed-symmetry tensor fields comes from the fact that $\bm{\dd}\ne\bm{\nabla}$, 
unlike scalars and $p$-forms. In attempting to covariantize the action \eqref{msgal} by replacing partial derivatives 
by covariant ones, this will play a crucial role. Since the general case is rather involved, let us focus on the special case of 
$p=q$. We comment on the general case in the concluding section. 

First, let us consider the simplest possible case of the
$(1,1)$ tensor field $h$, which was already found to correspond to a linearised graviton for any Lovelock term. The 4D case is 
rather simple to handle. The candidate covariant action is 
\be 
S_{2}[h, g]= S_{\text{LL}}[g]+\int\dd^4 x\int\dd^4\theta\, \dd^4\chi\,
\sqrt{-g} \,
\bm{g}\,\bm{\nabla}\bm{h}\, 
\widetilde{\bm{\nabla}}{\bm{h}}~.
\ee
This action has evidently only second order field equations, and therefore no dangerous higher derivatives appear either for the linearised
graviton $h$ or for the external graviton $g$. However, a more careful analysis reveals that such a theory is not consistent \cite{Aragone:1979bm}.  The main reason is that the divergence of the field equations for this action does not vanish. This leads to the problematic feature that the corresponding equation is only satisfied by a vanishing field $h_{ij}$.
This is also in accord with the general theorem of \cite{Boulanger:2000rq}, stating that under general assumptions there is no 
consistent coupling of two massless gravitons for which the free field limit is a non-interacting sum of massless Fierz-Pauli actions. This clearly shows that 
the requirement of second order field equations is a necessary but not sufficient one in order to consistently couple two fields. 

Despite these consistency issues, it is interesting and meaningful to examine whether non-linear actions containing higher 
derivatives can be covariantized. If for example one considers pure Gauss-Bonnet gravity in five 
dimensions, the linearised theory has nothing to do with the massless Fierz-Pauli action and thus it does not fall within the 
realm of the theorem of Ref. \cite{Boulanger:2000rq}. One could thus ask whether it is possible to couple two massless 
gravitons in Gauss-Bonnet gravity. 
 Secondly, 
already the requirement of second order field equations for such higher derivative theories is highly non-trivial and even more so 
in the mixed-symmetry case. Indeed, exactly because 
$\bm{\nabla}\bm{\o}^{(p,q)}\ne\bm{\dd}\bm{\o}^{(p,q)}$ for mixed-symmetry tensors, many more dangerous terms will appear as compared to $p$-forms. Compensating these dangerous terms is highly non-trivial. 

Let us therefore pursue the covariantization of the action \eqref{gbflat}, as motivated by the above arguments. First we state the net result.
The action 
\bea
\label{addcovactionp1}
S_{3}[h, g]=  S_{\text{LL}}[g]+ \int\dd^5 x\int\dd^5\theta\, \dd^5\chi\, 
\sqrt{-g}\, \left(\bm{\nabla}\bm{h}\, 
\widetilde{\bm{\nabla}}{\bm{h}}\, \bm{\nabla}\widetilde{\bm{\nabla}}\bm{h}+\widetilde{\bm{\nabla}}{\bm{h}}\, 
\widetilde{\bm h}_{l} \;\bm{H}^l \; \bm{Riem}\right)~,
\eea
where $\widetilde {\bm{h}}_{l}=h_{li}\theta^i$, $\bm{H}_{l}=H_{lij}\theta^i\chi^j$ and $H_{lij}=\sfrac 34 \nabla_lh_{ij}-\nabla_{(i}h_{j)l}$, has second order field equations, 
both with respect to $h$ and to $g$. (Indices here are raised and lowered using $g$.) We remind once more that the 
covariant derivative is taken with the metric $g$. Thus its action on $h$ introduces a potential interaction among the two.

Let us present some steps of the calculation in order to highlight the differences to the scalar and 
$p$-form Galileons. Although the calculation can be performed directly in the graded formalism, it is 
more instructive to follow it in components. Thus, in order to avoid carrying along quantities that play no role, we first define the shorthand notation
\be 
\int_{\mc{M}} :=\int \dd^5 x\int\dd^5\theta\, \dd^5\chi\, \theta^{i_1} \theta^{i_2} \theta^{i_3} \theta^{i_4} \theta^{i_5}
\chi^{j_1}\chi^{j_2}\chi^{j_3}\chi^{j_4}\chi^{j_5} \, \sqrt{-g} ~. 
\ee 
The main operations that may be freely performed are (a) exchange of all dummy indices $i \leftrightarrow j$ 
and (b) a minus sign for any exchange of two $i$ or two $j$ indices. Then our action is 
\bea
S_3 &=&
S_{\text{LL}}[g] 
-\int_{\mc{M}}\nabla_{i_1}h_{i_2j_2}\nabla_{j_1}h_{i_3j_3}\nabla_{i_4}\nabla_{j_4}h_{i_5j_5}+\int_{\mc{M}} 
H^l_{\ i_1j_1}\nabla_{j_2}h_{i_{2}j_{3}}h_{li_3}R_{i_4i_5j_4j_5}
\nn\\
&:=& S_{\text{LL}}[g]  + S^{(1)}[g,h]+S^{(2)}[g,h]~.
 \label{addcovactionp2}
\eea 
The second term is the covariantization of the corresponding flat space-time action, while the third
term is a counter-term or compensator, responsible for canceling third order  derivatives on $h$ and $g$. (There are no fourth order derivatives.)
First we examine the variation with respect to the field $h$. Since we are not interested in the second order terms, we present only the third order ones; this is denoted by a $\sim$ sign, as in Ref. \cite{covariantgalileons}. Using the basic formula 
\be 
[\nabla_{i_1},\nabla_{i_2}]h_{ij}=-R^l_{\ ii_1i_2}h_{lj}-R^{l}_{\ ji_1i_2}h_{li}=-R^{l}_{\ ji_1i_2}h_{li}~,
\ee 
due to the first Bianchi identity $R_{i[jkl]}=0$, and the 
symmetries of the Riemann tensor, we find 
\be 
\delta_{h}S^{(1)}\sim -{3} \int_{\mc{M}}\d h_{i_2j_3} \nabla_{i_1}h_{i_3j_1}\nabla_{i_4}R^l_{\ i_5j_2j_4}h_{lj_5}~.
\ee 
On the other hand, using repeatedly the second Bianchi identity $\nabla_{[i}R_{jk]lm}=0$ and once more the symmetries 
of the Riemann tensor, we compute 
\be 
\d_h S^{(2)}\sim -\delta_{h}S^{(1)}~.
\ee 
Thus the variation with respect to $h$ is free of higher derivatives.

Now we turn to the variation with respect to the metric $g$. Here we encounter the main difference to the scalar and $p$-form case, in that 
\be 
\d_g(\nabla_ih_{jk})=-\d_g\G^l_{ij}h_{lk}-\d_g\G^l_{ik}h_{lj}~. 
\ee
It is largely due to this fact that both third derivatives in the field $h$ as well as third derivatives of the metric $g$ will appear in the variations. For example we first find that
\be 
\d_gS^{(1)}\sim \d_g\G^l_{i_2j_2}\nabla_{i_1}h_{i_3j_1}\left[h_{lj_3}\left(2\nabla_{i_4}\nabla_{j_4}+\nabla_{j_4}\nabla_{i_4}\right)
-\sfrac 12 h_{li_4}[\nabla_{j_4},\nabla_{j_3}]\right]
h_{i_5j_5}~.
\ee 
Note that $\d_g\G^l_{ij}=g^{ll'}\left(\partial_i(\d g_{l'j})+\partial_j(\d g_{l'i})-\partial_{l'}(\d g_{ij})\right)$ and by partial integration third derivatives will appear in $\d_gS^{(1)}$. 
Similar considerations hold for $\d_g S^{(2)}$. The full calculation is tedious, but in order to understand the 
structure and the non-triviality of the cancellations, 
let us schematically express the variation of the full action as 
  \be 
  \d_g S_3=\int_{\mc{M}}\d g_{ij} \times K^{ij} +\int_{\mc{M}} \d g_{il} \times L^{il}~,
  \ee
  where $K$ and $L$ are long expressions containing third derivatives of $g$ (both) and third derivatives 
  of $h$ (only $K$).\footnote{A term with $\d g_{jl}$ can be absorbed in $L$ by an $i\leftrightarrow j$ exchange, 
  while a potential $\d g_{lm}$ term vanishes identically due to the second Bianchi identity.}  
 Now let us take a step back and return for a moment to the $h$ variation. The two terms cancelled 
 by virtue of the factor $3/4$ in the first term of $H_{lij}$. It turns out that the same factor is 
 responsible for the vanishing of the third derivatives on $h$ in the term $K$. Thus what remains is 
 the third derivatives of $g$ in both $K$ and $L$. These have to cancel separately because there is 
 no way to relate these two variations, due to their index structure. In fact $L$ is easier and it 
 indeed cancels out by means of the chosen coefficient in front of the second (symmetrized) term in $H_{lij}$. 
 This leaves no more tunable factors and the term involving $K$ constitutes a non-trivial test. 
 Fortunately, a long calculation establishes that indeed all terms within $K$ cancel out as well. 
 Thus, indeed the action \eqref{addcovactionp1}, or equivalently \eqref{addcovactionp2}, leads to 
 second order field equations. We further comment on this result and on the general case in the concluding section.
 
\section{Discussion and Conclusions}

Galileons for scalar fields were introduced in the spirit of avoiding
higher than second order derivatives in the field equations.
In flat space-time, the structure of such actions has been clarified in recent years
and also generalized in several directions, e.g.~for Abelian $p$-forms.

In this paper we introduced an alternative formulation for such theories, using two sets of graded, anticommuting variables.
This formalism serves two purposes. One is that all known Galileons are expressed in an elegant, index-free form, 
with clear properties and a simple way to test the instances when they become trivial - just by examining whether there is 
a square of an oddly graded variable involved. 

Interestingly, our approach reveals yet another generalization, in that it works also for mixed-symmetry tensor fields 
of type $(p,q)$. This is 
possible because such fields have two sets of antisymmetrized indices, which can be attributed to each of the two sets of anticommuting 
Grassmann variables - or, in the standard formulation, distributed evenly among the two epsilon tensors.
In addition, we find that in the simplest case of a $(1,1)$ tensor field $h$, the corresponding Galileon-type action is identified with 
linearised gravity, in particular with linearised Lovelock terms in diverse dimensions. In fact \emph{all} bosonic actions that are of importance in physics can be elegantly expressed in the graded formalism introduced in this paper.  

The central property of second order field equations is ensured by the generalization of the Galileon symmetry, which for mixed-symmetry  tensors is given in \eqref{gen-Gal}. Remarkably, for the gravity case, these are simply linearised diffeomorphisms, underlining the close similarity between Galileon scalars and Lovelock gravity. The natural gauge-invariant field strength for a mixed symmetry tensor is   given by the $(p+1,q+1)$ tensor $\bm{\dd} \widetilde{\bm {\dd}} {\bm{\omega}}$. Upon Hodge dualization, this yields
 \begin{align} {\rm field:~} & \quad (p,q) \rightarrow (D-p-2, D-q-2) \,, \notag \\
{\rm field~strength:~} & \quad (p+1,q+1) \rightarrow (D-p-1, D-q-1) \,. 
 \end{align}
 This is the bi-form extension of the usual Hodge duality between forms (of which a shift-symmetric scalar is the special case with $p=0$). As a consequence, a scalar with Galileon invariance can similarly be described by a $(D-2,D-2)$ bi-form. Similarly, linearised gravity  is dual to $(D-3,D-3)$. Note that $D=4$ gravity is clearly a special case, allowing for self-dual gravity.

An intriguing question is whether Galileon-type theories for mixed-symmetry tensor fields can be covariantized, as happens for their 
scalar and $p$-form counterparts. 
In general, when one covariantizes the flat Galileon action for a field, variation with respect to the field 
generically leads to 
higher than second order derivatives acting on the metric,
while the field itself is safe from being acted upon by third or higher 
derivatives.
On the other hand, when the action is varied with respect to the metric,
third derivatives on the field are encountered, and
 depending on the type of field one is working with, third derivatives on the metric itself occur too. 
 The notable feature in the scalar and $p$-form case
is that the latter issue does not appear,
effectively due to the use of the Levi-Civita connection.
Thus, in those cases
the equation of motion for the field suffers from 
third order derivatives on the metric---but not on the field, while the equation of motion for the metric 
suffers from third derivatives on the field---but not on the metric.
Elimination of such dangerous terms is achieved by the introduction of compensating 
actions with explicit curvature dependence.

For covariant mixed-symmetry tensor fields the situation changes, due to the fact that 
their covariant derivative contains {Christoffel symbols}, unlike the scalar and $p$-form case. The consequence is then  that 
the variation with respect to the metric leads to third derivatives on the metric itself. This is a new feature, and 
it makes the task of determining compensating actions much harder. However, in this paper we showed that in the case of 
$(1,1)$ field, thrice appearing in a five-dimensional action, a single simple compensator is enough to cancel 
simultaneously the third derivatives on the field and on the metric. In fact, this seems possible for any $(p,p)$ field, a 
special case of $(p,q)$ for $p=q$, while the $p\ne q$ case is in principle analogous but this is less clear at the moment. 

We find this result remarkable, 
even though it does not guarantee the full consistency of the theory, but only the absence of an Ostrogradsky ghost. 
There are certainly reasons why such a theory might end up having consistency issues. For example its four-dimensional 
analogs are doomed to either freeze the field or violate unitarity \cite{Aragone:1979bm}. Recalling that such 
four-dimensional theories should lead to the massless Fierz-Pauli action in the free-field limit, the general theorem of 
Ref. \cite{Boulanger:2000rq} applies, stating that interactions between massless gravitons are impossible. Whether this remains the 
case in the presence of higher-derivatives, for example in the five-dimensional theory we discussed, is left for further study. 

\paragraph{Acknowledgments.} We would like to thank C. Deffayet, R. Klein and D. Stefanyszyn for useful discussions. This work was supported in part by COST Action MP1405 ``Quantum Structure of Spacetime''. A. Chatzistavrakidis acknowledges partial support by the Croatian Science Foundation under the project IP-2014-09-3258.
F. S. Khoo and P. Schupp acknowledge support by the DFG Research Training Group 1620 ``Models of Gravity''. F. S. Khoo thanks the Van Swinderen Institute for hospitality during part of this project.

\appendix 

\section{Computational details}

\subsection{Scalars in flat space-time}

Let us first provide some more details in order to justify the statements of Section 2. We begin with the equivalence of the action 
\eqref{ggscalar} to the standard scalar Galileon action \eqref{pigal}.
Using the definitions from Section \ref{gggal1}, the action \eqref{ggscalar} becomes
\bea 
S^{\text{Gal,1}}_{n+1}[\pi]&=&\frac 1{(D-n)!}\int\dd^D x\int \dd^D\theta \, \dd^D  \chi \, (\eta_{kl}\theta^k\chi^l)^{D-n}(\theta^{i_1}\partial_{i_1}\pi) 
  (\chi^{j_1}\partial_{j_1}\pi)(\theta^i\chi^j\partial_i\partial_j\pi)^{n-1} 
  \nn\\
  &=& \frac 1{(D-n)!}\int\dd^D x\int \dd^D\theta \, \dd^D  \chi \, \eta_{k_{n+1}l_{n+1}}\dots \eta_{k_Dl_D}\theta^{k_{n+1}}\dots\theta^{k_D} 
	  \chi^{l_{n+1}}\dots\chi^{l_D}\times 
	  \nn\\
	  && \,\,\,\,\,\,\,\,\,\,\,\,\,\,\,\,\,\,\,\,\, \,\,\,\,\,\,\,\,\, \,\,\,\,\, \,\,\,\,\, \,\,\, \times\,
	    \theta^{i_1}\dots\theta^{i_{n}}\chi^{j_1}\dots\chi^{j_{n}}\pi_{i_1}\pi_{j_1} 
	    \pi_{i_2j_2}\dots\pi_{i_{n}j_{n}} \nn\\ 
	    &=&\frac 1{(D-n)!}\int\dd^D x\int \dd^D\theta \, \dd^D  \chi \, \theta^1\dots\theta^D\chi^1\dots\chi^D 
	    \vare^{i_1\dots i_{n}k_{n+1}\dots k_D}\vare^{j_1\dots j_{n}l_{n+1}\dots l_D}\times   \nn\\
	      &&\,\,\,\,\,\,\,\,\,\,\,\,\,\,\,\,\,\,\,\,\, \,\,\,\,\,\,\,\,\, \,\,\,\,\, \,\,\,\,\, \,\,\, \times \,
	      \eta_{k_{n+1}l_{n+1}}\dots\eta_{k_Dl_D}	\pi_{i_1}\pi_{j_1} 
	    \pi_{i_2j_2}\dots\pi_{i_{n}j_{n}}  
	    \nn\\
	    &=&\frac 1{(D-n)!}\int\dd^D x \, \vare^{i_1\dots i_{n}k_{n+1}\dots k_D}\vare^{j_1\dots j_{n}}_{\ \ \ \ \ \ k_{n+1}\dots k_D} 
	    \pi_{i_1}\pi_{j_1} \pi_{i_2j_2}\dots\pi_{i_{n}j_{n}}~,\label{proofscalar}
\eea
which is indeed the desired action \eqref{pigal} in the notation $\pi_i=\partial_i\pi$ and $\pi_{ij}=\partial_i\partial_j\pi$. For this derivation we used the following two properties:
\bea 
&& \theta^{i_1}\dots\theta^{i_D}=\vare^{i_1\dots i_D}\theta^1\dots\theta^D~,
 \qquad  \int \dd^D\theta \, \dd^D  \chi \, \theta^1\dots\theta^D\chi^1\dots\chi^D=1~,
\eea
and we recall here that $$\vare^{i_1\dots i_D}=\d^{i_1\dots i_D}_{1\dots D} \quad  \text{and} \quad 
\vare_{i_1\dots i_D}= 
\d^{1\dots D}_{i_1\dots i_D}$$ are the Levi-Civita symbols. All other actions are obtained in a similar fashion, after integrating out the 
graded variables.

The property \eqref{dda} is easily proven as follows:
\bea 
(\bm{\dd}\widetilde{\bm{\dd}}\widetilde{\bm{A}})^2&=&(\theta^i\chi^j\chi^k\partial_i\partial_jA_k)(\theta^l\chi^m\chi^n 
\partial_l\partial_mA_n)\nn\\
&=&\theta^i\theta^l\chi^j\chi^k\chi^m\chi^n \partial_i\partial_jA_k\partial_l\partial_mA_{n} 
\nn\\
&=&\sfrac 12\theta^i\theta^l\chi^j\chi^k\chi^m\chi^n \partial_i\partial_jA_k\partial_l\partial_mA_n  
+\sfrac 12 \theta^l\theta^i\chi^m\chi^n\chi^j\chi^k \partial_l\partial_mA_n\partial_i\partial_jA_k
\nn\\
&=& \sfrac 12(\theta^i\theta^l+\theta^l\theta^i)\chi^j\chi^k\chi^m\chi^n \partial_i\partial_jA_k\partial_l\partial_mA_n =0~,
\eea 
and similarly for $\bm{A}$. 

For completeness, let us mention that  \eqref{pigal} is not the only possible expression for  scalar Galileon actions.  Alternative expressions are known \cite{reviewgalileons} and in the graded notation take for example the following form:
\bea \label{gal2}
S^{\text{Gal,2}}_{n+1}[\pi]&=& \frac 1{(D-n+1)!}\int\dd^Dx\int d^D\theta \, d^D  \chi \, \bm{\eta}^{D-n+1} \,
\partial_i \pi \,\widetilde{\bm{\dd}}(\partial^i \pi) \,\bm{\dd}\pi\, (\bm{\dd}\widetilde{\bm{\dd}} \pi )^{n-2}~,
\\ \label{gal3}
S^{\text{Gal,3}}_{n+1}[\pi]&=&\frac 1{(D-n+1)!}\int\dd^Dx\int d^D\theta \, d^D  \chi \, \bm{\eta}^{D-n+1} \,
\partial_i \pi \,\partial^i \pi \, (\bm{\dd}\widetilde{\bm{\dd}} \pi )^{n-1}~.
\eea
The appearance of contracted terms in these expressions can be understood as arising from the fact that  $\bm{\eta}^{D-n} \, \widetilde{\bm{\dd}}\pi$ is related to the Hodge-dual of  $\bm{\dd} \pi$.

\subsection{Scalars in curved space-time}\label{sec.covscaform}

For completeness, let us also review  the scalar case in curved space-time.
In the graded formalism, the natural would-be covariantization of the scalar Galileon is 
\be 
S_{n+1}[\pi,g] =S_{\text{EH}} [g]-\frac 1{(D-n)!} \int\dd^Dx\int d^D\theta \, d^D  \chi \, \sqrt{-g}\, \bm{g}^{D-n} \,
\pi  \, (\bm{\nabla}\widetilde{\bm{\nabla}} \pi )^{n}~,
\ee
where the first term is the Einstein-Hilbert action. Varying with respect to the scalar field $\pi$, one encounters the following 
non-vanishing, higher-than-second-order terms:
\bea 
\widetilde{\bm{\nabla}}\pi  (\bm{\nabla}\widetilde{\bm{\nabla}} \pi )^{n-2} \bm{\nabla}^2\widetilde{\bm{\nabla}} \pi~, \quad 
 \bm{\nabla}\pi (\bm{\nabla}\widetilde{\bm{\nabla}} \pi )^{n-2} \widetilde{\bm{\nabla}}\bm{\nabla}\widetilde{\bm{\nabla}} \pi~,\quad
\bm{\nabla}\pi\widetilde{\bm{\nabla}}\pi  (\bm{\nabla}\widetilde{\bm{\nabla}} \pi )^{n-3}
\widetilde{\bm{\nabla}}\bm{\nabla}^2\widetilde{\bm{\nabla}} \pi~.
\eea
The first two do not spoil the property of second order field equations, since they correspond to 
a curvature tensor (second derivatives on $g$)  multiplying a covariant scalar field (first derivative on $\pi$). However, the third term indeed induces higher-derivative terms in the field equation. 
As suggested in \cite{generalizedgalileons,covariantgalileons}, one can remedy this by adding 
compensating terms in the action in order to eliminate such higher-derivative 
contributions to the field equations. In the present formalism this is simply implemented via the couplings
\be 
S_{n+1,r}[\pi,g] = \frac{1}{(D-n)!} \int\dd^Dx\int d^D\theta \, d^D  \chi \, \sqrt{-g}\, \bm{g}^{D-n} \,
\bm{\nabla} \pi \, \widetilde{\bm{\nabla}}\pi \, (\bm{\nabla}\widetilde{\bm{\nabla}} \pi )^{n-2r-1}(\nabla_i\pi\nabla^i\pi\bm{Riem})^r~.
\ee
Then, as proven in \cite{generalizedgalileons}, the action 
\be 
S[\pi,g]=\sum_{r=0}^{\lfloor \sfrac {n-1}2\rfloor} {\cal C}_{(n+1,r)}S_{n+1,r}[\pi,g]~,
\ee
with $S_{n+1,0}:=S_{n+1}$, 
has second order field equations for both $\pi$ and $g$. The coefficients are given as 
\be 
{\cal C}_{(n+1,r)}=\left(-\frac 18\right)^r\frac{(n-1)!}{(n-1-2r)!(r!)^2}~,
\ee
so that in particular ${\cal C}_{(n+1,0)}=1$.

\subsection{Two-forms in curved space-time}

Let us now review the case of a $2$-form in curved space-time, 
proven in \cite{pformgalileons}.
The covariantized action at 7 dimensions is
\bea
\label{cov2form}
S_4[B^{(2,0)},g] =
 \int\dd^7x\int d^7\theta \, d^7  \chi \, \sqrt{-g}\;
 \bm\nabla \bm B \; \tn\widetilde{\bm B} \; \bm\nabla\tn\widetilde{\bm B}
 \; \tn \bm\nabla \bm B~.
\eea
When Levi-Civita connection is assumed, which consequently means $\bm\nabla\bm B = \bm\dd \bm B$,\footnote{A normalized totally antisymmetric $3$-form,
$\bm H := \bm\dd\bm B$, or equivalently $\widetilde{\bm H} := \td\widetilde{\bm B}$ can be defined.}
the action becomes
\bea
S_4[B^{(2,0)},g] =
 \int\dd^7x\int d^7\theta \, d^7  \chi \, \sqrt{-g}\;
 \bm\dd \bm B \; \td\widetilde{\bm B} \; \bm\nabla\td\widetilde{\bm B}
 \; \tn \bm\dd \bm B~.
 \label{Bcovac}
\eea
The only higher-than-second-order term resulting from the variation of the action (\ref{Bcovac}) with respect to the $2$-form tensor field is
$
2\, \bm\dd \bm B \; \td\widetilde{\bm B} \; \td\bm\nabla\tn\bm\dd\bm B,
$ which contains four derivatives, that is,
third derivatives on $g$.
One remedies this by adding to the action (\ref{Bcovac}) the compensating term 
\bea
S_{4,1}[B^{(2,0)},g] = - {\frac{9}{4} }
 \int\dd^7x\int d^7\theta \, d^7  \chi \, \sqrt{-g}\;
  \bm H \; \widetilde{\bm H} \; 
  \widetilde{\bm H}^{\ell} \bm{H}_{\ell}\;
  \bm{Riem}~,
\eea
where
$\bm {H}_{\ell} = H_{\ell i_1 i_2} \theta^{i_1}\theta^{i_2}$ and
$\widetilde{\bm H}_{\ell} = H_{\ell j_1 j_2}\chi^{j_1} \chi^{j_2} $.

\end{document}